\documentclass[pra,twocolumn,showpacs]{revtex4-1}
\usepackage{amsmath,amscd,amsfonts,amssymb}
\usepackage{graphicx,amsfonts}
\usepackage[pdftex]{hyperref}
\begin{document}
\title{Evolution of tripartite entangled states in a
decohering environment and their experimental protection using
dynamical decoupling}
\author{Harpreet Singh}
\email{harpreetsingh@iisermohali.ac.in}
\affiliation{Department of Physical Sciences, Indian
Institute of Science Education \& 
Research (IISER) Mohali, Sector 81 SAS Nagar, 
Punjab 140306 India.}
\author{Arvind}
\email{arvind@iisermohali.ac.in}
\affiliation{Department of Physical Sciences, Indian
Institute of Science Education \& 
Research (IISER) Mohali, Sector 81 SAS Nagar, 
Punjab 140306 India.}
\author{Kavita Dorai}
\email{kavita@iisermohali.ac.in}
\affiliation{Department of Physical Sciences, Indian
Institute of Science Education \& 
Research (IISER) Mohali, Sector 81 SAS Nagar, 
Punjab 140306 India.}
\begin{abstract}
We embarked upon the task of experimental protection of
different classes of tripartite entangled states, namely the
maximally entangled GHZ and W states and the ${\rm W
\bar{W}}$ state, using dynamical decoupling.  The states
were created on a three-qubit NMR quantum information
processor and allowed to evolve in the naturally noisy NMR
environment.  Tripartite entanglement was monitored at each
time instant during state evolution, using negativity as an
entanglement measure.  It was found that the W state is most
robust while the GHZ-type states are most fragile against
the natural decoherence present in the NMR system.  The
${\rm W \bar{W}}$ state which is in the GHZ-class, yet
stores entanglement in a manner akin to the W state,
surprisingly turned out to be more robust than the GHZ
state.  The experimental data were best modeled by
considering the main noise channel to be an uncorrelated
phase damping channel acting independently on each qubit,
alongwith a generalized amplitude damping channel.  Using
dynamical decoupling, we were able to achieve a significant
protection of entanglement for GHZ states.  There was a
marginal improvement in the state fidelity for the W state
(which is already robust against natural system
decoherence), while the ${\rm W \bar{W}}$ state showed a
significant improvement in fidelity and protection against
decoherence.
\end{abstract}
\maketitle
\section{Introduction}
\label{intro}
Quantum entanglement is considered to lie at the crux of
QIP~\cite{nielsen-book} and while two-qubit entanglement can be completely
characterized, multipartite entanglement is more difficult to quantify and is
the subject of much recent research~\cite{horodecki-rmp-09}.  Entanglement can
be rather fragile under decoherence and various multiparty entangled states
behave very differently under the same decohering
channel~\cite{dur-prl-04}.  It is hence of paramount importance to
understand and control the dynamics of multipartite entangled states in 
multivarious noisy
environments~\cite{mintert-pr-05,aolita-prl-08,aolita-rpp-15}.

A three-qubit system is a good model system to study the diverse response of
multipartite entangled states to decoherence and  the
entanglement dynamics of three-qubit GHZ and W states were  theoretically
studied~\cite{borras-pra-09,weinstein-pra-10}.
Under an arbitrary (Markovian) decohering environment, it was shown that W
states are more robust than GHZ states for certain kinds of channels while the
reverse is true for other kinds of
channels~\cite{carvalho-prl-04,siomau-eurphysd-10,siomau-pra-10,ali-jpb-14}.

On the experimental front, tripartite entanglement was generated using 
photonic qubits and the robustness of W state entanglement was
studied in optical 
systems~\cite{lanyon-njp-09,zang-scirep-15,he-qip-15,zang-optics-16}.  The 
dynamics of multi-qubit entanglement under
the influence of decoherence was experimentally characterized using a string of
trapped ions~\cite{barreiro-nature} and in
superconducting qubits~\cite{wu-qip-16}. 
In the context of NMR quantum information
processing, three-qubit entangled states were experimentally
prepared~\cite{suter-3qubit,shruti-generic,manu-pra-14}, and their decay rates
compared with bipartite entangled states~\cite{kawamura-ijqc-06}.

With a view to protecting entanglement,
dynamical decoupling (DD) schemes have been successfully
applied to decouple a multiqubit system from both transverse
dephasing and longitudinal relaxation
baths~\cite{viola-review,uhrig-njp-08,kuo-jmp-12,zhen-pra-16}.  UDD
schemes have been used in the context of entanglement
preservation~\cite{song-ijqi-13,franco-prb-14}, and it was
shown theoretically that Uhrig DD schemes are able to
preserve the entanglement of two-qubit Bell states and
three-qubit GHZ states for quite long
times~\cite{agarwal-scripta}.

In this work, we experimentally explored the robustness
against decoherence, of three different tripartite entangled
states, namely, the GHZ, W and $\rm W{\bar W}$ states.  The
${\rm W{\bar W}}$ state is a novel tripartite entangled
state which belongs to the GHZ entanglement class in the
sense that it is SLOCC equivalent to the GHZ state, however
stores its entanglement in ways very similar to that of the W 
state~\cite{Devi2012,shruti-wwbar}.  We 
created these states with a very high fidelity, via 
GRAPE-optimized rf pulses~\cite{tosner-jmr-09}
on a system of three NMR qubits, using three fluorine
spins individually addressable in frequency space.
We
allowed these entangled states to decohere and 
measured their
entanglement content at different instances in time.  
To estimate the fidelity of state preparation
and entanglement content, we performed
complete state 
tomography~\cite{leskowitz-pra-04} using maximum
likelihood estimation~\cite{singh-pla-16}.  
As a measure for
tripartite entanglement, we used a well-known extension of the
bipartite Peres-Horodecki separability
criterion~\cite{peres-prl-96} called
negativity~\cite{vidal-pra-02}.

Our results showed that the W state was most robust against the environmental
noise, followed by the ${\rm W {\bar W}}$ state, while the GHZ state was rather
fragile.  We analytically solved the Lindblad master equation for decohering
open quantum systems and showed that the best-fit 
to our experimental data was provided by a model  
which considered two predominant noise channels
acting on the three qubits: 
and a homogeneous phase-damping channel acting independently on 
each qubit 
and a generalized amplitude damping channel.
Next, we protected entanglement of these states using
two different DD sequences: the symmetrized XY-16(s) and the Knill dynamical
decoupling (KDD) sequences, and evaluated their efficacy of protection. Both DD
schemes were able to achieve a good degree of entanglement protection.  The GHZ
state was dramatically protected, with its entanglement persisting for nearly
double the time. The W state showed a marginal improvement,
which was to be expected since these DD schemes are designed to protect mainly
against dephasing noise, and our results indicated that the W state 
is already robust against this type of decohering channel.
Interestingly, although the $\rm W {\bar W}$ state belongs to the GHZ
entanglement-class, our experiments revealed that its entanglement
persists for a longer time than the GHZ state, while
the DD schemes 
are able to preserve its entanglement to a reasonable
extent.  The decoherence characteristics of the ${\rm W \bar{W}}$ state hence
suggest a way of protecting fragile GHZ-type 
states against noise by transforming the type of
entanglement (since a GHZ-class state can be transformed via local operations
to a ${\rm W\bar{W}}$ state). 
These aspects of the entanglement
dynamics of the ${\rm W {\bar W}}$ state require more
detailed studies for a better understanding.

There has been a longstanding debate about the existence of entanglement in
spin ensembles at high temperature as encountered in NMR experiments. There are
two ways to look at the situation. Entangled states in such ensembles are
obtained via unitary transformations on pseudopure states.  If we consider the
entire spin ensemble, given that the number of spins that are involved in the
pseudopure state is very small compared to the total number of spins, it has
been shown that the overall ensemble is not entangled~\cite{braunstein,chuang}.
However one can take a different point of view and only consider the
sub-ensemble of spins that have been prepared in the pseudopure state, and as
far as these spins are concerned, entanglement genuinely
exists~\cite{brazil1,brazil2,long}. The states that we have created are
entangled in this sense, and hence may not be considered as entangled if one
works with the entire ensemble. Therefore, one has to be aware and cautious
about this aspect while dealing with these states.  These states are sometimes
referred to as being pseudo-entangled.  Moreover, these states have interesting
properties in terms of the presence of multiple-quantum coherences and their
evolution and dynamics under decoherence.

This paper is organized as follows:~In
Section~\ref{entang-decoh} we describe the experimental
decoherence behaviour of tripartite entangled states, with
section~\ref{system} containing details of the NMR system
section~\ref{construct} delineating the experimental schemes
to prepare tripartite-entangled GHZ, W and $\rm W{\bar W}$
states.  The experimental entanglement dynamics of these
states decohering in a noisy environment  is contained in
section~\ref{decay}.  Section~\ref{ddprotect} describes the
results of protecting these tripartite entangled states
using robust dynamical decoupling sequences, while
Section~\ref{concl} presents some conclusions.  The
theoretical model of noise damping used to fit the
experimental data is described in the Appendix.  
\section{Dynamics of tripartite entangled states}
\label{entang-decoh}
\subsection{Three-qubit NMR system}
\label{system}
\begin{figure}[h]
\centering
\includegraphics[angle=0,scale=1.0]{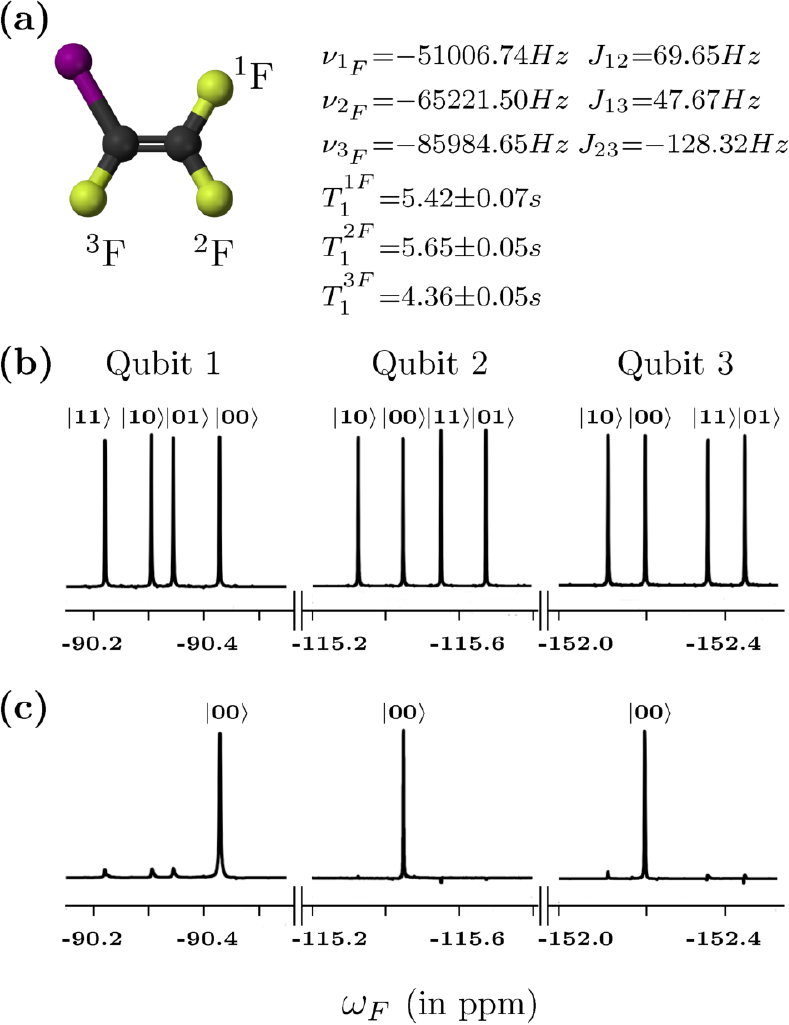}
\caption{(a) Molecular structure of trifluoroiodoethylene molecule and
tabulated system parameters  with chemical shifts $\nu_i$
and scalar couplings J$_{ij}$ (in Hz), and spin-lattice
relaxation times $T_{1}$ and spin-spin relaxation times
T$_{2}$ (in seconds).  (b) NMR spectrum obtained after a
$\pi/2$ readout pulse on the thermal equilibrium state.  and
(c) NMR spectrum of the pseudopure $\vert 000 \rangle$
state.  The resonance lines of each qubit are labeled by the
corresponding logical states of the other qubit.
}
\label{molecule}
\end{figure}
We use the
three ${}^{19}$F nuclear spins of the 
trifluoroiodoethylene (C$_2$F$_3$I) molecule
to encode the three qubits. On an NMR
spectrometer operating at 600 MHz, the fluorine
spin resonates at a Larmor frequency of 
$\approx 564$ MHz.
The molecular structure of the three-qubit system with
tabulated system parameters and the NMR spectra of the
qubits at thermal equilibrium and prepared in the
pseudopure state $\vert 000 \rangle$ are shown in
Figs.~\ref{molecule}(a), (b), and (c), respectively.
The Hamiltonian of a 
weakly-coupled three-spin system in a frame 
rotating at $\omega_{{\rm rf}}$ (the frequency of the
electromagnetic field $B_1(t)$ applied to manipulate spins
in a static magnetic field $B_0$)
is given by~\cite{ernst-book-87}:
\begin{equation}
{\cal H} = -\sum_{i=1}^3 (\omega_i -
\omega_{{\rm rf}}) I_{iz} 
+ \sum_{i<j,j=1}^3 2 \pi J_{ij} I_{iz} I_{jz}
\end{equation}
where $I_{iz}$ is the spin angular momentum
operator in the $z$ direction for ${}^{19}$F; the
first term in the Hamiltonian denotes the 
Zeeman interaction between the fluorine spins and the
static magnetic field $B_0$ with $\omega_i = 2 \pi \nu_i$ being
the Larmor frequencies; the second term represents
the spin-spin interaction  with $J_{ij}$ being the 
scalar coupling constants.
The three-qubit equilibrium density matrix (in the high
temperature and high field approximations) is in a highly
mixed state given by:
\begin{eqnarray}
\rho_{eq}&=&\tfrac{1} {8}(I+\epsilon \ \Delta \rho_{eq})
\nonumber \\
\Delta\rho_{{\rm eq}} &\propto& \sum_{i=1}^{3} I_{iz}
\end{eqnarray}
with a thermal polarization $\epsilon \sim 10^{-5}$, $I$
being
the $8 \times 8$ identity operator and 
$\Delta \rho_{{\rm eq}}$ being the deviation part of
the density matrix.
The system was first initialized into the $\vert 000\rangle$
pseudopure state using the spatial averaging
technique~\cite{cory-physicad},
with the density operator given by
\begin{equation}
\rho_{000}=\frac{1-\epsilon}{8}I 
+ \epsilon \vert 000\rangle\langle000 \vert
\end{equation}

\begin{figure}[h]
\centering
\includegraphics[angle=0,scale=1.0]{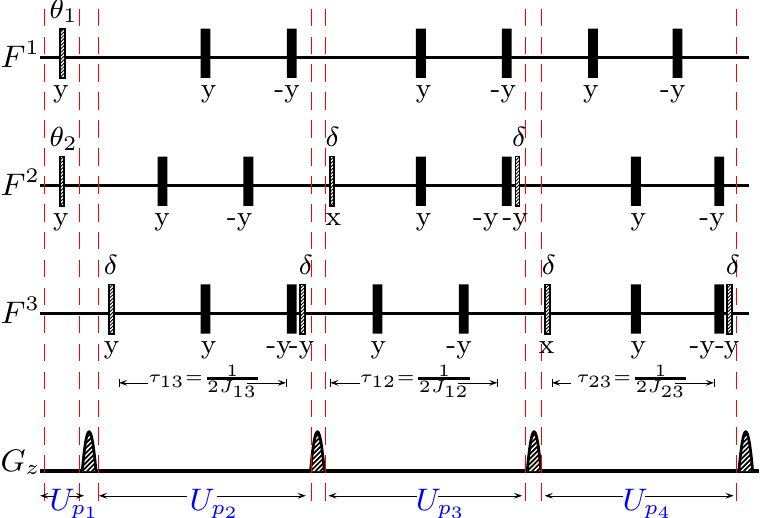}
\caption{NMR pulse sequence used to prepare
pseudopure state $\rho_{000}$ starting from 
thermal equilibrium.The
pulses represented by black filled rectangles are of angle
$\pi$. The other rf flip angles are set to $\theta_1=\frac{5\pi}{12}$, 
$\theta_2=\frac{\pi}{6}$ and $\delta=\frac{\pi}{4}$. The phase of
each rf pulse is written below each pulse bar. The evolution
interval $\tau_{ij}$ is set to a multiple of the scalar
coupling strength ($J_{ij}$).
}
\label{ppure-fig}
\end{figure}

The specific sequence of rf pulses, $z$ gradient pulses and
time evolution periods we used to 
prepare the pseudopure state
$\rho_{000}$ starting from thermal equilibrium is shown in
Figure~\ref{ppure-fig}.
All the rf pulses used
in the pseudopure state preparation scheme were constructed
using the Gradient Ascent Pulse Engineering (GRAPE)
technique~\cite{tosner-jmr-09} and were designed to be
robust against rf inhomogeneity, with an average fidelity of
$ \ge 0.99$. Wherever possible, two independent
spin-selective rf pulses were combined using a specially
crafted single GRAPE pulse;  for instance the first two rf
pulses to be applied before the first field gradient pulse,
were combined into a single pulse specially crafted pulse
($U_{p_{1}}$ in Figure~\ref{ppure-fig}), of duration
$600 \mu$s. The combined pulses $U_{p_{2}}$, $U_{p_{3}}$
and $U_{p_{4}}$ applied later in the
sequence were of a total duration $\approx 20$ ms.

All experimental density matrices were reconstructed using a
reduced tomographic protocol and by using maximum likelihood
estimation~\cite{leskowitz-pra-04,singh-pla-16} with the set
of operations $\{ III, IIY, IYY, YII, XYX, XXY, XXX\}$; $I$
is the identity (do-nothing operation) and $X (Y)$ denotes a
single spin operator implemented by a spin-selective $\pi/2$
pulse.  We constructed these spin-selective pulses for
tomography using GRAPE, with the length of each pulse
$\approx 600 \mu$s.  The fidelity of an experimental density
matrix was estimated by measuring the projection between the
theoretically expected and experimentally measured states
using the Uhlmann${\rm -}$Jozsa fidelity
measure~\cite{uhlmann-fidelity,jozsa-fidelity}:
\begin{equation}
F =
\left(Tr \left( \sqrt{
\sqrt{\rho_{\rm theory}}
\rho_{\rm expt} \sqrt{\rho_{\rm theory}}
}
\right)\right)^2
\label{fidelity}
\end{equation}
where $\rho_{\rm theory}$ and $\rho_{\rm expt}$ denote the
theoretical and experimental density matrices
respectively.
The experimental density matrices were reconstructed by
repeating each experiment ten times (keeping the temperature fixed at 288
K). The mean of the ten experimentally reconstructed 
density matrices was used to compute the
statistical error in the state fidelity. 
The experimentally created
pseudopure state $\vert 000\rangle$ was tomographed with 
a fidelity of $0.985 \pm 0.015$ and the total time taken to prepare the
state was $\approx 60$ ms.
\begin{figure}[h]
\centering
\includegraphics[angle=0,scale=1.0]{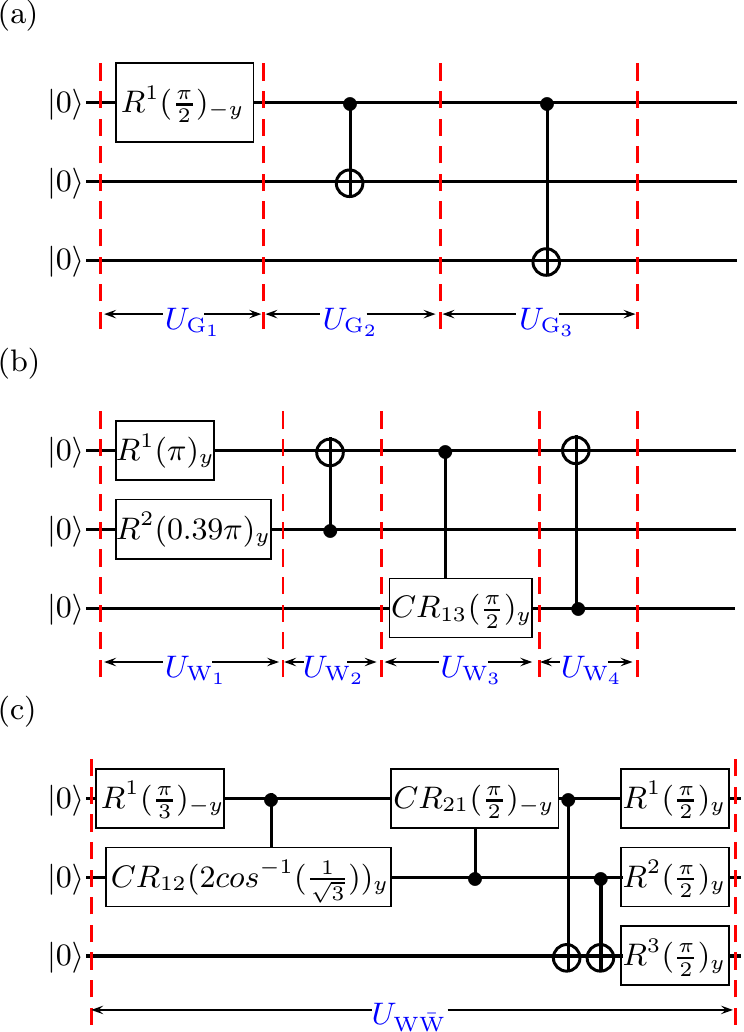}
\caption{(Quantum circuit showing 
the sequence of implementation of the
single-qubit local rotation gates (labeled by 
$R$), two-qubit controlled-rotation gates (labeled
by $CR$) and controlled-NOT
gates required
to construct the (a) GHZ state (b) W state and (c) 
${\rm W\bar{W}}$ state.}
\label{ckt}
\end{figure}
\subsection{NMR implementation of tripartite entangled states}
\label{construct}
Tripartite entanglement has been well characterized and it
is known that the two different classes of tripartite
entanglement, namely GHZ-class and W-class, are
inequivalent. While both classes are maximally entangled,
there are differences in the their
type of entanglement: the W-class entanglement is more
robust against particle loss than the GHZ-class (which
becomes separable if one particle is lost) and it is
also known that the W state has the maximum possible
bipartite entanglement in its reduced two-qubit 
states~\cite{guhne-review}. The entanglement in the
${\rm W \bar{W}}$ state (which belongs to the GHZ-class
of entanglement) shows a surprising result, that it is
reconstructible from its reduced two-qubit states (similar
to the W-class of states).

We now turn to the construction of tripartite entangled
states on the three-qubit NMR system. The quantum circuits
to prepare the three qubits in a GHZ-type state, a W state
and a ${\rm W \bar{W}}$ state are shown in Figs.~\ref{ckt}
(a), (b) and (c), respectively. Several of the quantum gates
in these circuits were optimized using the GRAPE algorithm
and we were able to achieve a high gate fidelity and
smaller pulse lengths.

The GHZ-type
$\frac{1}{\sqrt{2}}(\vert000\rangle-\vert111\rangle)$ state
was prepared from the $\vert000\rangle$ pseudopure state by
a sequence of three quantum gates (labeled as $U_{G_1},
U_{G_2}, U_{G_3}$ in Fig.~\ref{ckt}(a)): first a selective
rotation of $\left [\frac{\pi}{2}\right]_{-y}$ on the first
qubit, followed by a CNOT$_{12}$ gate, and finally a
CNOT$_{13}$ gate.  The step-by-step sequential gate
operation leads to: 
\begin{eqnarray}
\vert 0 0 0 \rangle &\stackrel{{R^1{\left
(\frac{\pi}{2}\right)_{-y}}}}{\longrightarrow}&
\frac{1}{\sqrt{2}}\left(\vert 0 0 0 \rangle -
\vert 1 0 0 \rangle \right) \nonumber \\
&\stackrel{\rm CNOT_{12}}{\longrightarrow}&
\frac{1}{\sqrt{2}}\left(\vert 0 0 0 \rangle -
\vert 1 1 0 \rangle\right) \nonumber \\
&\stackrel{\rm CNOT_{13}}{\longrightarrow}&
\frac{1}{\sqrt{2}}\left(\vert 0 0 0 \rangle -
\vert 1 1 1 \rangle \right)
\end{eqnarray}
All the pulses for the three gates used for GHZ state
construction were designed using the GRAPE algorithm and had
a fidelity $\ge$ 0.995. The GRAPE pulse duration
corresponding to the gate $U_{G_1}$ is $600 \mu $s, while
the $U_{G_2}$ and $U_{G_3}$ gates had pulse durations of
$24$ms. The GHZ-type state was prepared with a fidelity of
$0.969 \pm 0.013$.
\begin{figure}[h]
\centering
\includegraphics[angle=0,scale=1.0]{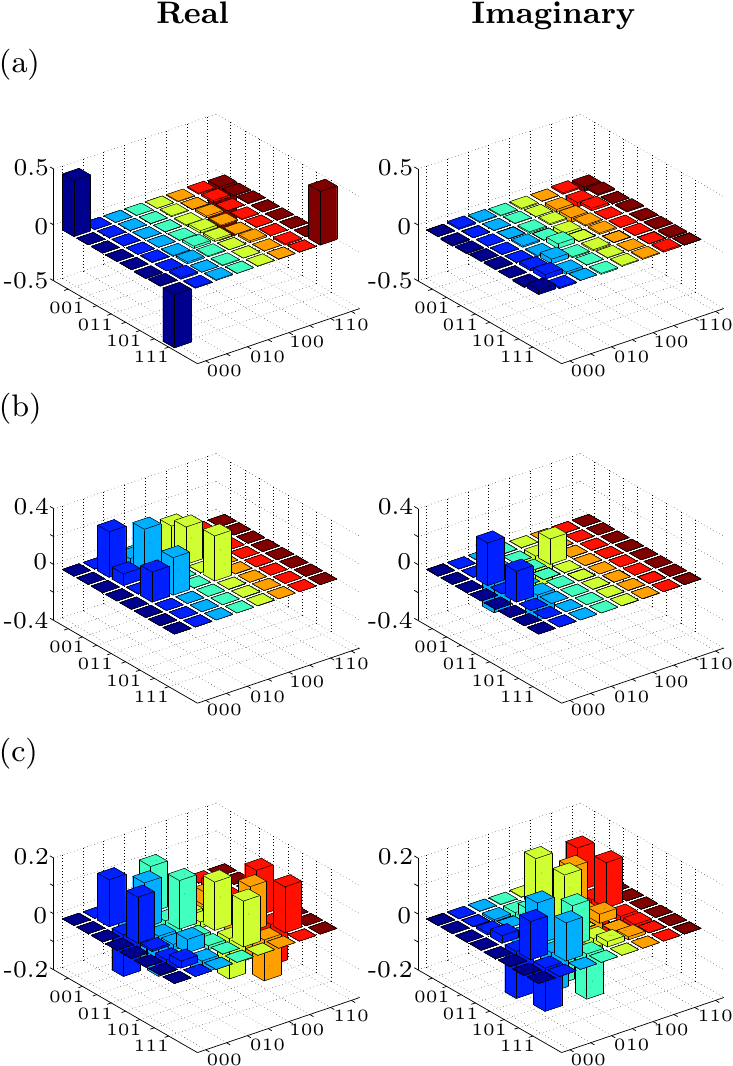}
\caption{The real (left) and imaginary (right) parts of the
experimentally tomographed (a) GHZ-type state, with
a fidelity of $0.969 \pm 0.013$. (b)  
W state, with a fidelity of $0.964 \pm 0.012$ and (c)  
$W\bar{W}$ state with a fidelity of $0.937 \pm 0.005$.
The rows and columns encode the
computational basis in binary order from
$\vert 000 \rangle$ to $\vert 111 \rangle$.}
\label{nodd}
\end{figure}

The W state was prepared from the initial $\vert 000 \rangle$ by a
sequence of four unitary operations (labeled as $U_{W_1}, U_{W_2}, 
U_{W_3}, U_{W_4}$ in Fig.~\ref{ckt}(b)) and the sequential
gate operation leads to:
\begin{eqnarray}
\vert 000 \rangle &
\stackrel{R^{1}\left({\pi}\right)_y}{\longrightarrow} &
\vert 100 \rangle \nonumber \\
 & \stackrel{\rm R^{2}\left({0.39\pi}\right)_y}
{\longrightarrow} &
\sqrt{\frac{2}{3}}
\vert 100 \rangle + \frac{1}{\sqrt{3}} \vert 110 \rangle
\nonumber \\
& \stackrel{\rm CNOT_{21}} {\longrightarrow} &
\sqrt{\frac{2}{3}} \vert 100
\rangle +
\frac{1}{\sqrt{3}}  \vert 010 \rangle
\nonumber \\
& \stackrel{\rm
CR_{13}{\left(\frac{\pi}{2}\right)_y}}{\longrightarrow} &
\frac{1}{\sqrt{3}} [\vert 100 \rangle + 
\vert 101 \rangle + \vert 010 \rangle]
\nonumber \\
& \stackrel{\rm CNOT_{31}} {\longrightarrow} &
\frac{1}{\sqrt{3}}[
\vert 100 \rangle +  \vert
001 \rangle +
\vert 010 \rangle] 
\label{weqn}
\end{eqnarray}
The different unitaries were individually optimized using
GRAPE and the  pulse duration for $U_{W_1}$, $U_{W_2}$, $U_{W_3}$, and
$U_{W_4}$ turned out to be $600 \mu $s, $24 $ms, $16 $ms, and $20 $ms,
respectively and the fidelity of the final state was
estimated to be $0.937 \pm 0.012$.

The ${\rm W\bar{W}}$ state was constructed by
applying the following sequence of gate operations on  
the $\vert000\rangle$ state:
\begin{eqnarray}
\vert 000 \rangle &
\stackrel{R^{1}\left(\frac{\pi}{3}\right)_{-y}}{\longrightarrow}
&
\frac{\sqrt{3}}{2}\vert 000 \rangle-\frac{1}{2}\vert 100
\rangle \nonumber \\
 & \stackrel{\scriptstyle {\rm
CR}_{12}\left(0.61\pi\right)_y}
{\longrightarrow} &
\frac{\sqrt{3}}{2}\vert 0 0 0 \rangle -
\frac{1}{2\sqrt{3}} \vert 1 0 0 \rangle -
\sqrt{\frac{1}{6}} \vert 1 1 0 \rangle \nonumber \\
& \stackrel{\scriptstyle {\rm
CR}_{21}\left(\frac{\pi}{2}\right)_{-y}}
{\longrightarrow} &
\frac{1}{2}(\sqrt{3}\vert 0 0 0 \rangle -
\frac{1}{\sqrt{3}} (\vert 1 0 0 \rangle +
\vert 1 1 0 \rangle +  \nonumber \\
& & \vert 0 1 0 \rangle))    \nonumber \\
& \stackrel{\rm CNOT_{13}}{\longrightarrow} &  \frac{1}{2}(
\sqrt{3}\vert 0 0 0 \rangle -
\frac{1}{\sqrt{3}} (\vert 1 0 1 \rangle
+ \vert 1 1 1 \rangle + \nonumber \\
& &\vert 0 1 0 \rangle)) \nonumber \\
& \stackrel{\rm CNOT_{23}}{\longrightarrow} &  \frac{1}{2}(
\sqrt{3}\vert 0 0 0 \rangle -
\frac{1}{\sqrt{3}} (\vert 1 0 1 \rangle
+ \vert 1 1 0 \rangle + \nonumber \\
& & \vert 0 1 1 \rangle))\nonumber \\
& \stackrel{\rm
R^{123}\left(\frac{\pi}{2}\right)_y}{\longrightarrow} &
\frac{1}{\sqrt{6}}(
\vert 0 0 1 \rangle + \vert 0 1 0 \rangle +
\vert 0 1 1 \rangle + \nonumber \\
& &  \vert 1 0 0 \rangle + \vert 1 0 1 \rangle + \vert 1 1 0
\rangle)
\label{wwbareqn}
\end{eqnarray}
The unitary operator for the entire preparation
sequence (labeled $U_{W\bar{W}}$ in Fig.~\ref{ckt}(c))
comprising a spin-selective rotation operator:~two
controlled-rotation gates, two controlled-NOT gates
and one non-selective rotation by $\frac{\pi}{2}$ on all
the three qubits,
was created by a specially crafted single
GRAPE pulse (of pulse length $48$ms) and applied to the initial state
$\vert000\rangle$. The final state had a computed fidelity of
$0.937 \pm 0.005$.
\subsection{Decay of tripartite entanglement}
\label{decay}
\begin{figure}[h]
\centering
\includegraphics[angle=0,scale=1.0]{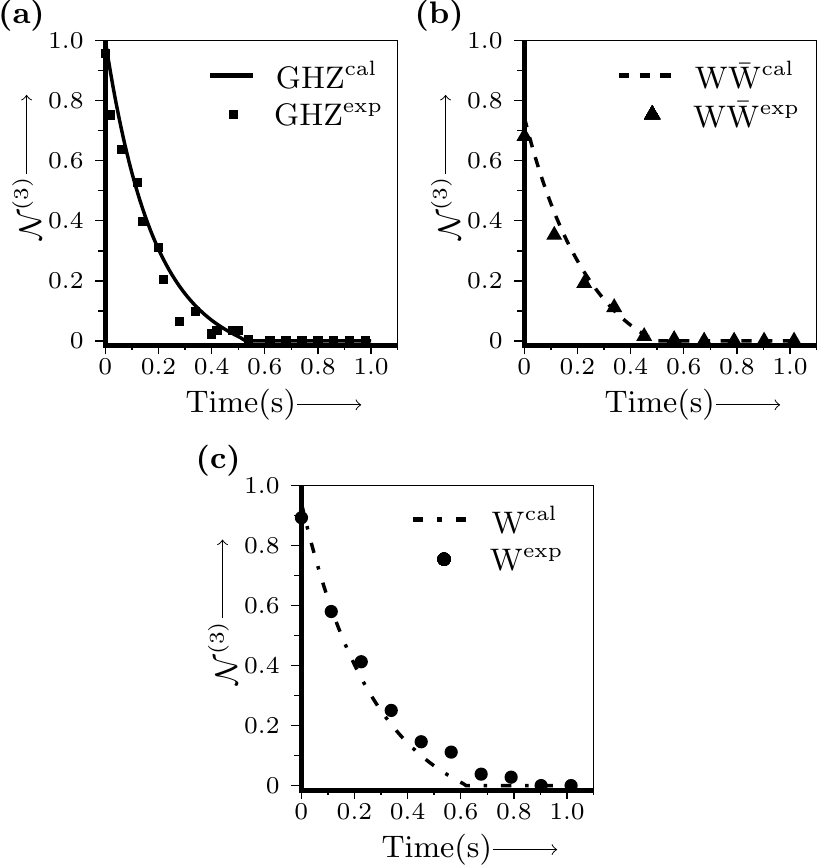}
\caption{Time dependence of the tripartite negativity
${\cal N}^{(3)}$ for the three-qubit system initially
experimentally 
prepared in the (a) GHZ state (squares) (b) W state (circles) 
and (c) ${\rm W {\bar W}}$
state (triangles) (the superscript ${\rm exp}$ denotes
``experimental data''). 
The fits are the calculated decay of negativity ${\cal N}^{(3)}$ of
the GHZ state (solid line), the $W\bar{W}$ state (dashed line) and 
the W state (dotted-dashed line), 
under the action of the modeled NMR noise 
channel (the superscript ${\rm cal}$ denotes ``calculated fit''). 
The W state is most robust against the NMR noise
channel, whereas
the GHZ state is most fragile.}
\label{3qdecay}
\end{figure}

\begin{figure}[h]
\centering
\includegraphics[angle=0,scale=1.0]{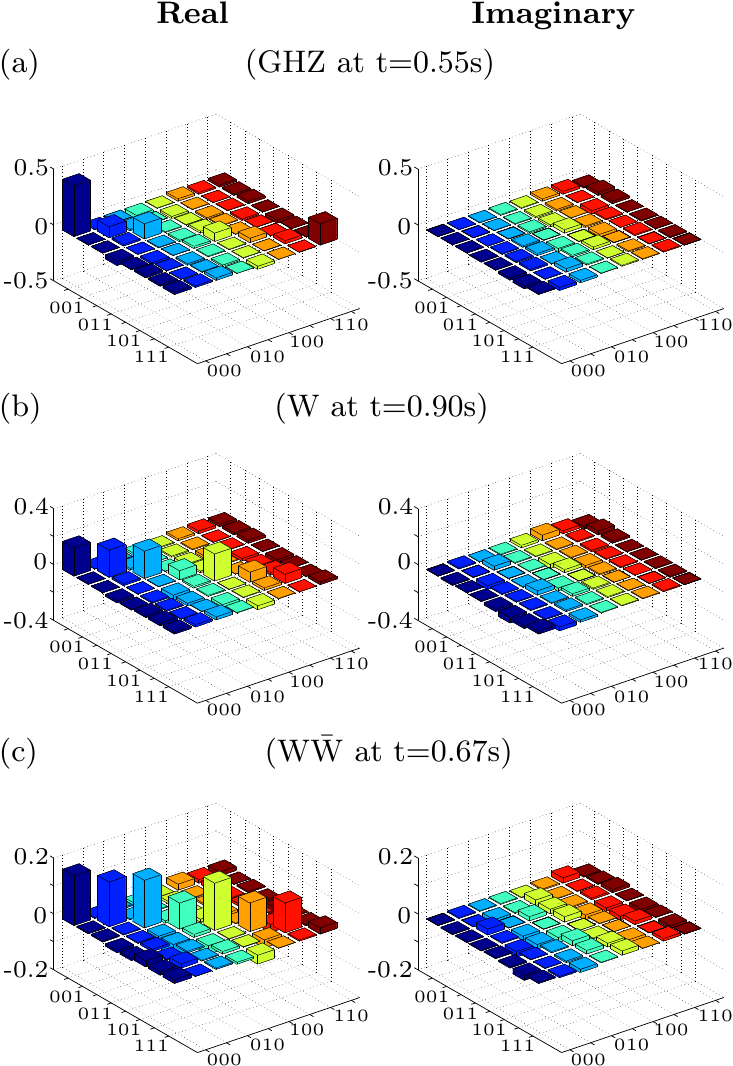}
\caption{The real (left) and imaginary (right) parts of the
the experimentally tomographed 
density matrix of the state at the
time instances when 
the tripartite negativity ${\cal N}^{(3)}_{123}$ approaches
zero for the (a) GHZ state at $t=0.55$ s (b) 
W state at $t=0.90$ s and (c) 
${\rm W\bar{W}}$ state at $t=0.67$ s.
The rows and columns encode the
computational basis in binary order, from
$\vert 000 \rangle$ to $\vert 111 \rangle$.}
\label{tomodecay}
\end{figure}

We next turn to the dynamics of tripartite entanglement under decoherence
channels acting on the system.  For two qubits, all entangled states are
negative under partial transpose (NPT) and for such NPT states, the minimum
eigenvalues of the partially transposed density operator is a measure of
entanglement~\cite{peres-prl-96}.  This idea has been extended to three qubits,
and entanglement can be quantified for our three-qubit system using the
well-known tripartite negativity ${\cal N}^{(3)}_{123}$
measure~\cite{vidal-pra-02,weinstein-pra-10}:
\begin{equation}
{\cal{N}}^{(3)}_{123}= [{\cal N}_{1}{\cal
N}_{2}{\cal N}_{3}]^{1/3} 
\end{equation}
where the negativity of a qubit ${\cal N}_i$ 
refers to the most negative eigenvalue of
the partial transpose of the density matrix with respect to the qubit $i$.  We
studied the time evolution of the tripartite negativity ${\cal N}^{(3)}_{123}$
for the tripartite entangled states, as computed from the experimentally
reconstructed density matrices at each time instant.  The experimental results
are depicted in Fig~\ref{3qdecay} (a), (b) and (c) for the GHZ state, the ${\rm
W \bar{W}}$ state, and the W state, respectively.  Of the three entangled
states considered in this study, the GHZ and W states are maximally entangled
and hence contain the most amount of tripartite negativity, while the ${\rm W
\bar{W}}$ state is not maximally entangled and hence has a lower tripartite
negativity value.  The experimentally prepared GHZ state initially has a ${\cal
N}^{(3)}_{123}$ of 0.96 (quite close to its theoretically expected value of
1.0).  The GHZ state decays rapidly, with its negativity approaching zero in
0.55 s.  The experimentally prepared ${\rm W \bar{W}}$ state initially has a
${\cal N}^{(3)}_{123}$ of 0.68 (close to its theoretically expected value of
0.74), with its negativity approaching zero at 0.67 s.  The experimentally
prepared W state initially has a ${\cal N}^{(3)}_{123}$ of 0.90 (quite close to
its theoretically expected value of 0.94).  The W state is quite long-lived,
with its entanglement persisting up to 0.9 s.  The tomographs of the
experimentally reconstructed density matrices of the GHZ, W and ${\rm W
\bar{W}}$ states at the time instances when the tripartite negativity parameter
${\cal N}^{(3)}_{123}$ approaches zero for each state, are displayed in
Fig.~\ref{tomodecay}.

We explored
the noise channels acting on our three-qubit
NMR entangled states which best fit our experimental data,
by analytically solving a master equation in
the Lindblad form, along the lines suggested
in Reference~\cite{jung-pra-08}.
The master equation is given by~\cite{lindblad}:
\begin{equation}
\frac{\partial \rho}{\partial t} 
= -i[H_s,\rho] + \sum_{i,\alpha}
\left[
L_{i,\alpha} \rho L_{i,\alpha}^{\dagger}
- \frac{1}{2} \{ L^{\dagger}_{i,\alpha} 
L_{i,\alpha},\rho
\}
\right]
\label{mastereqn} 
\end{equation}
where $H_s$ is the system Hamiltonian, 
$L_{i,\alpha} \equiv \sqrt{\kappa_{i,\alpha}} 
\sigma^{(i)}_{\alpha}$ is the Lindblad operator
acting on the $i$th qubit and $\sigma^{(i)}_{\alpha}$
is the Pauli operator on the $i$th qubit,
$\alpha=x,y,z$; the constant
$\kappa_{i,\alpha}$ turns out to be the inverse of the
decoherence time.
We consider a decoherence model wherein a nuclear spin is acted on by two noise
channels namely a phase damping channel (described by the T$_2$ relaxation in
NMR) and a generalized amplitude damping  channel (described by the T$_1$
relaxation in NMR)~\cite{childs-pra-03}.  As the fluorine spins  in our
three-qubit system have widely differing chemical shifts, we assume that each
qubit interacts independently with its own environment.  The experimentally
determined T$_1$ NMR relaxation rates are T$_1^{1F}=5.42\pm 0.07$ s,
T$_1^{2F}=5.65\pm 0.05$ s and T$_1^{3F}=4.36\pm 0.05$ s, respectively.  The
T$_2$ relaxation rates were experimentally measured by first rotating the spin
magnetization into the transverse plane by a $90^{\circ}$ rf pulse followed by
a delay and fitting the resulting magnetization decay.  The experimentally
determined T$_2$ NMR relaxation rates are T$_2^{1F}=0.53\pm 0.02$s,
T$_2^{2F}=0.55\pm 0.02$ s,  and T$_2^{3F}=0.52\pm 0.02$ s, respectively.  We
solved the master equation (Eqn.~(\ref{mastereqn})) for the GHZ, W and ${\rm
W\bar{W}}$ states with the Lindblad operators $L_{i,x} \equiv
\sqrt{\frac{\kappa_{i,x}}{2}}\sigma^{(i)}_{x}$ and $L_{i,z} \equiv
\sqrt{\frac{\kappa_{i,z}}{2}} \sigma^{(i)}_{z}$, where
$\kappa_{i,x}=\frac{1}{T_1^i}$ and $\kappa_{i,z}=\frac{1}{T_2^i}$.  With this
model, the GHZ  state decays at the rate $\gamma^{al}_{GHZ}=6.33\pm 0.06
s^{-1}$, and its entanglement approaches zero in 0.53 s.  The $W\bar{W}$ state
decays at the rate  $\gamma^{al}_{W\bar{W}}=5.90 \pm 0.10  s^{-1}$, and its
entanglement approaches zero in 0.50 s.  The $W$ state decays at the rate
$\gamma^{al}_{W}=4.84\pm 0.07 s^{-1}$, and its entanglement approaches zero in
0.62 s.  We used the high-temperature approximation (T $\approx \infty$) to
model the noise (the experiments were performed at 288 K), and the results of
the analytical calculation and the experimental data match well, as shown in
Figure~\ref{3qdecay}.
\section{Protecting three-qubit entanglement via dynamical
decoupling}
\label{ddprotect}
As the tripartite entangled states under investigation 
are robust against noise to varying extents, we wanted to
discover if either the amount of entanglement in these
states could be protected or their entanglement could be
preserved for longer times, using dynamical decoupling (DD)
protection schemes. 
While DD sequences are effective
in decoupling system-environment interactions, often errors
in their implementation arise either due to errors in the
pulses or errors due to off-resonant driving~\cite{suter-review}.
Two approaches have been used to design robust DD sequences
which are impervious to pulse imperfections:
the first approach replaces the $\pi$ rotation
pulses with composite pulses inside the DD sequence, while
the second approach focuses on 
optimizing phases of the pulses in the DD sequence. In
this work, we use DD sequences 
that use pulses with phases applied along
different rotation axes: the XY-16(s) and the 
Knill Dynamical Decoupling (KDD)
schemes~\cite{souza-pra-12}.
\begin{figure}[h]
\centering
\includegraphics[angle=0,scale=1.0]{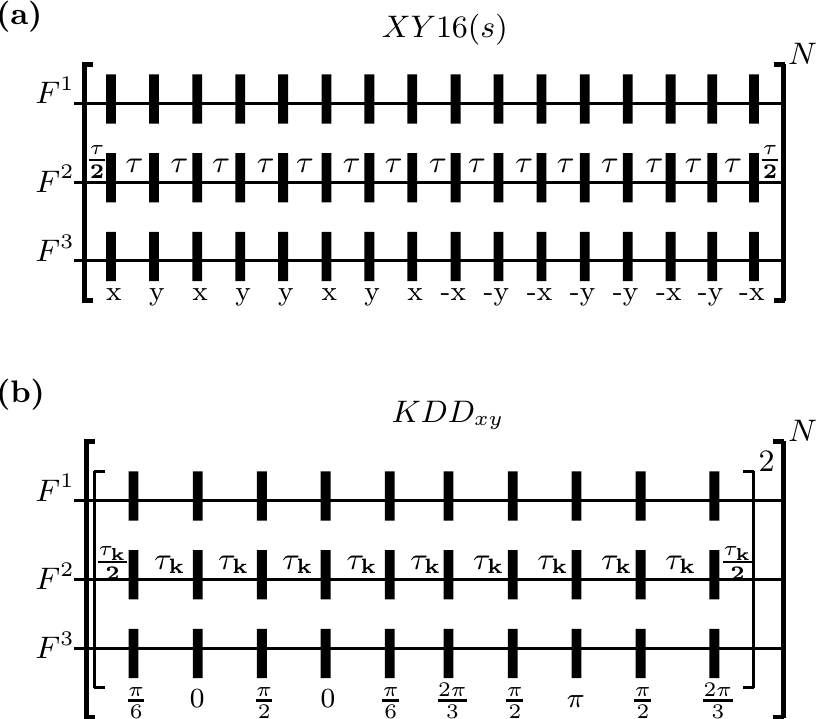}
\caption{NMR pulse sequence corresponding to (a) 
XY-16(s)
and (b) KDD$_{xy}$ DD schemes 
(the superscript 2 implies that  
the set of pulses inside the bracket is applied twice, to
form one cycle of the DD scheme).  The
pulses represented by black filled rectangles (in both
schemes) are of angle
$\pi$, and are applied simultaneously on all 
three qubits (denoted by $F^{i}, i=1,2,3$).
The angle below each pulse denotes the phase with which it
is applied. Each DD cycle is repeated $N$ times, with
$N$ large to achieve good system-bath decoupling.}
\label{dd-fig}
\end{figure}
In conventional DD schemes the $\pi$ pulses are applied
along one axis (typically $x$) and as a consequence, only
the coherence along that axis is well protected. The XY
family of DD schemes applies pulses along two perpendicular
($x,y$) axes, which protects coherence equally along both
these axes~\cite{souza-phil}.
The XY-16(s) sequence is 
constructed by combining an XY-8(s) cycle with
its phase-shifted copy,
where the (s) denotes the ``symmetric'' version i.e. 
the cycle is time-symmetric with respect to its center. The
XY-8 cycle is itself created by combining a 
basic XY-4
cycle with its time-reversed copy. One full unit cycle of
the XY-16(s) sequence comprises sixteen $\pi$ pulses
interspersed with free evolution time periods, and
each cycle
is repeated $N$ times for better decoupling.
The KDD sequence 
has  additional phases which further
symmetrize pulses in the $x-y$ plane and compensate
for pulse errors; each $\pi$ pulse
in a basic XY DD sequence is replaced by five $\pi$ pulses,
each of a different 
phase~\cite{ryan-prl-10,souza-prl-11}:
\begin{equation}
{\rm KDD}_{\phi} \equiv
(\pi)_{\frac{\pi}{6}+\phi}-
(\pi)_{\phi} -(\pi)_{\frac{\pi}{2}+\phi}-(\pi)_{\phi}
-(\pi)_{\frac{\pi}{6}+\phi}
\label{basickdd}
\end{equation}
where $\phi$ denotes the phase of the pulse; we set
$\phi=0$ in our experiments. The KDD$_{\phi}$ sequence of
five pulses given in Eqn.~\ref{basickdd} 
protects coherence along only one axis. To
protect coherences along both the $(x,y)$ axes, 
we use the KDD$_{xy}$ sequence, which combines two
basic five-pulse blocks shifted in phase by $\pi/2$ i.e
$[{\rm KDD}_{\phi} - {\rm KDD}_{\phi+\pi/2}]$.
One unit cycle of the KDD$_{xy}$ sequence contains two
of these pulse-blocks shifted in phase, for a total
of twenty $\pi$ pulses.
The XY-16(s) and KDD$_{xy}$ DD sequences  are given
in Figs.~\ref{dd-fig}(a) and (b) respectively, where the
black filled rectangles represent $\pi$ pulses on all three
qubits and $\tau$ ($\tau_k$) indicates a free evolution
time period.  We note here that the 
chemical shifts of the
three fluorine qubits 
in our particular
molecule cover a very large frequency
bandwidth, 
making it difficult to implement an accurate non-selective
pulse simultaneously on all  the qubits.  To circumvent this
problem, we crafted a special excitation pulse of duration $\approx
400 \mu$s  consisting of a set of three Gaussian shaped pulses
that are applied at different spin frequency offsets and are
frequency modulated to achieve simultaneous
excitation~\cite{shruti-wwbar}.
\begin{figure}[h]
\centering
\includegraphics[angle=0,scale=1.0]{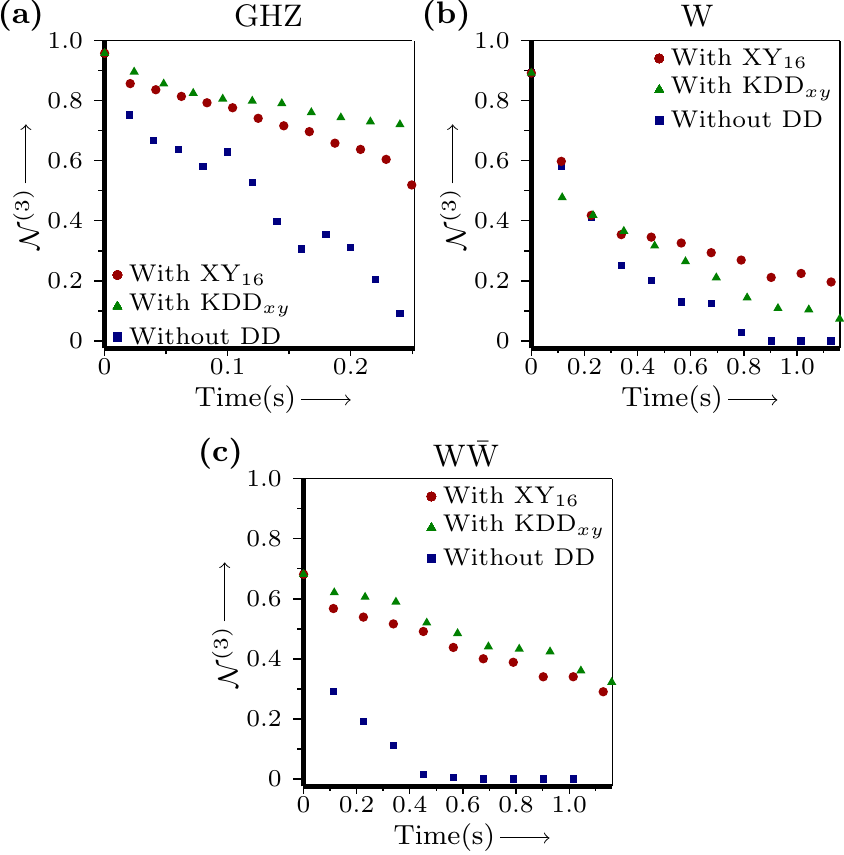}
\caption{Plot of the tripartite  negativity (${\cal N}^{(3)}_{123}$) with
time, computed for the (a) GHZ-type state, (b) W state
and (c) $W\bar{W}$ state. The negativity was computed for
each state without applying any protection and after
applying the XY-16(s) and KDD$_{xy}$ dynamical decoupling
sequences.Note that the time scale for part (a) is
different from (b) and (c)}
\label{entang-dd}
\end{figure}
Figs.\ref{entang-dd}(a),(b) and (c) show the results 
of protecting the GHZ, W and ${\rm W\bar{W}}$ states
respectively, using the XY-16(s) and the KDD$_{xy}$
DD sequences.  

\noindent{\bf GHZ state protection:}
The XY-16(s) protection scheme was implemented on the GHZ state with
an inter-pulse delay of $\tau = 0.25$ ms and one run of the
sequence took $10.40$ ms (including the length of
the sixteen $\pi$ pulses). The value of the negativity
${\cal N}^3_{123}$ 
remained close to 
0.80 and 0.52 for up to $80$ms and $240$ ms respectively
when XY-16 protection was applied, while 
for the unprotected state 
the state fidelity is
quite low and 
${\cal N}^3_{123}$ decayed  to 
a low value of 0.58 and 0.09  at $80$ms and $240$ ms, respectively
(Fig.~\ref{entang-dd}(a)).  The KDD$_{xy}$ protection scheme on this
state was implemented with an inter-pulse delay $\tau_k
=0.20$ ms and one run of the sequence took $12$ ms
(including the length of the twenty $\pi$ pulses).  The value of
the negativity ${\cal N}^3_{123}$ remained close to 0.80 and 0.72 for
up to $140$ms and $240$ ms when KDD$_{xy}$ protection was applied
(Fig.~\ref{entang-dd}(a)).

\noindent{\bf W state protection:}
The XY-16(s) protection scheme was implemented on the
W state with an
inter-pulse delay $\tau = 3.12$ ms 
and one run of the
sequence took $56.40$ ms (including the length of the
sixteen $\pi$ pulses). 
The value of the negativity ${\cal
N}^3_{123}$ remained close to 0.30 for up to $0.68$ s when XY-16
protection was applied, whereas ${\cal N}^3_{123}$ reduced 
to 0.1
at $0.68$ s 
 when no state protection is applied
(Fig.~\ref{entang-dd}(b)).
The KDD$_{xy}$ protection scheme was implemented
on the W state with an inter-pulse delay 
$\tau_k = 2.5$ ms and one run of the
sequence took $58$ ms (including the length of the
twenty $\pi$ pulses). 
The value of the negativity ${\cal
N}^3_{123}$ remained close to 0.21 for upto $0.70$ s when KDD$_{xy}$
protection was applied 
(Fig.~\ref{entang-dd}(b)).

\noindent{\bf ${\rm W \bar{W}}$ state protection:}
The XY-16(s) protection sequence was
implemented on the ${\rm W \bar{W}}$ state
with an inter-pulse delay of $\tau = 3.12$ ms and one run of the
sequence took $56.40$ ms (including the length of the
sixteen $\pi$ pulses). 
The value of the negativity ${\cal
N}^3_{123}$ remained close to  0.5 for upto $0.45$ s when
XY-16(s) protection was applied, whereas ${\cal N}^3_{123}$ 
reduced almost to zero ($\approx 0.02$)
at $0.45$ s 
when no protection was applied
(Fig.~\ref{entang-dd}(c)).
The KDD$_{xy}$ protection sequence was
applied with an inter-pulse delay of 
$\tau_k = 2.5$ ms and one run of the
sequence took $58$ ms (including the length of the
twenty $\pi$ pulses). 
The value of the negativity ${\cal
N}^3_{123}$ remained close to 0.52 for upto $0.46$ s when KDD$_{xy}$
protection was applied
(Fig.~\ref{entang-dd}(c)).

The results of UDD-type of protection summarized
above demonstrate that state protection worked to varying
degrees and protected the entanglement of the tripartite
entangled states to different extents, depending on the
type of state to be protected. The GHZ state showed maximum
protection and the ${\rm W\bar{W}}$ state also showed
a significant amount of protection, while the W state
showed a marginal improvement under
protection.
We note here that the lifetime of
the GHZ state is not significantly enhanced by using
DD state protection; what is noteworthy is that state
fidelity remains high (close to 0.8) under DD
protection, whereas the
state quickly gets disentangled (fidelity drops to 0.4) when
no protection is applied. This implies that under DD
protection, there is no leakage from the state to other
states in the Hilbert space of the three qubits. 
\section{Conclusions}
\label{concl}
We undertook an experimental study of the 
dynamics of tripartite entangled states in a three-qubit
NMR system. Our results are relevant in the
context of other studies which
showed that different entangled states exhibit varying
degrees of robustness against diverse noise channels. We
found that the W state was the most robust against
the decoherence channel acting on the three NMR qubits, the
GHZ state was the most fragile and decayed very quickly,
while the ${\rm W \bar{W}}$ state was more robust than the
GHZ state but less robust than the W state. We also
implemented entanglement protection on these states using
dynamical decoupling sequences. The protection worked to a
remarkable extent in entanglement preservation in the GHZ
and ${\rm W \bar{W}}$ states, while the W state showed a
better fidelity under protection but no appreciable increase
in the lifetime of entanglement. 
The entangled states that we deal with in this study 
are obtained by unitary transformations
on pseudopure states, where only a small subset of spins
participate, and are thus pseudo-entangled.
Our results have important
implications for entanglement storage and preservation in
realistic quantum information processing protocols.

\begin{acknowledgments}
All experiments were performed on a Bruker Avance-III 600
MHz FT-NMR spectrometer at the NMR Research Facility at
IISER Mohali.  
KD acknowledges funding from DST India under
Grant No. EMR/2015/000556.
Arvind acknowledges funding from DST India under
Grant No. EMR/2014/000297.
HS acknowledges CSIR India for financial support.
\end{acknowledgments}
\raggedbottom
\appendix*
\section{Analytical solution of the Lindblad master equation}
\label{anal}
We analytically solved a master equation of the
Lindblad form given in
Eqn.~\ref{mastereqn}~\cite{lindblad,jung-pra-08},
by putting in explicit values for the Lindblad
operators according to the two main NMR noise
channels (generalized amplitude damping and phase
damping), and computed the decay behavior of the
GHZ, W and ${\rm W \bar{W}}$ states.  

Under the simultaneous action of all the NMR noise
channels, the GHZ state decoheres as:

\begin{equation} 
\rho_{GHZ}=\left( \begin{array}{cccccccc} \alpha_1  & 0 & 0
& 0 & 0 & 0 & 0 & \beta_1 \\ 0 & \alpha_2  & 0 & 0 & 0 & 0 & \beta_2 & 0 \\ 0 &
0 & \alpha_3  & 0 & 0 & \beta_3 & 0 & 0 \\ 0 & 0 & 0 & \alpha_4  & \beta_4 & 0
& 0 & 0 \\ 0 & 0 & 0 & \beta_4 & \alpha_4  & 0 & 0 & 0 \\ 0 & 0 & \beta_3 & 0 &
0 & \alpha_3  & 0 & 0 \\ 0 & \beta_2 & 0 & 0 & 0 & 0 & \alpha_2  & 0 \\ \beta_1
& 0 & 0 & 0 & 0 & 0 & 0 & \alpha_1 \\ \end{array} \right) 
\end{equation}
where 
\begin{eqnarray} 
\alpha_1 &=&
\frac{1}{8}(1+e^{-(\kappa_{x,1}+\kappa_{x,2})t}+e^{-(\kappa_{x,1}+\kappa_{x,3})t}
+e^{-(\kappa_{x,2}+\kappa_{x,3})t}) 
\nonumber \\ \alpha_2 &=&
\frac{1}{8}(1+e^{-(\kappa_{x,1}+\kappa_{x,2})t}-e^{-(\kappa_{x,1}+\kappa_{x,3})t}
-e^{-(\kappa_{x,2}+\kappa_{x,3})t}) \nonumber \\ \alpha_3 &=&
\frac{1}{8}(1-e^{-(\kappa_{x,1}+\kappa_{x,2})t}+e^{-(\kappa_{x,1}+\kappa_{x,3})t}
-e^{-(\kappa_{x,2}+\kappa_{x,3})t}) \nonumber \\ 
\alpha_4 &=&
\frac{1}{8}(1-e^{-(\kappa_{x,1}+\kappa_{x,2})t}-e^{-(\kappa_{x,1}+\kappa_{x,3})t}
+e^{-(\kappa_{x,2}+\kappa_{x,3})t}) \nonumber \\ 
\beta_1 &=&
\frac{1}{8}(e^{-(\kappa_{1,x}+\kappa_{2,x}+\kappa_{3,x}+\kappa_{1,z}+\kappa_{2,z}+\kappa_{3,z})t}
\nonumber \\ &&
\quad(e^{\kappa_{1,x}t}+ e^{\kappa_{2,x}t}+e^{\kappa_{3,x}t}
+e^{(\kappa_{1,x}+\kappa_{2,x}+\kappa_{3,x})t}) \nonumber \\
\beta_2 &=&
\frac{1}{8}(e^{-(\kappa_{1,x}+\kappa_{2,x}+\kappa_{3,x}+\kappa_{1,z}+\kappa_{2,z}+\kappa_{3,z})t}
\nonumber \\ &&
\quad (-e^{\kappa_{1,x}t}-e^{\kappa_{2,x}t}+e^{\kappa_{3,x}t}
+e^{(\kappa_{1,x}+\kappa_{2,x}+\kappa_{3,x})t}) \nonumber \\
\beta_3 &=&
\frac{1}{8}(e^{-(\kappa_{1,x}+\kappa_{2,x}+\kappa_{3,x}+\kappa_{1,z}+\kappa_{2,z}+\kappa_{3,z})t}
\nonumber \\ 
\quad &&(-e^{\kappa_{1,x}t}+ e^{\kappa_{2,x}t}-e^{\kappa_{3,x}t}
+e^{(\kappa_{1,x}+\kappa_{2,x}+\kappa_{3,x})t}) \nonumber \\
\beta_4 &=&
\frac{1}{8}(e^{-(\kappa_{1,x}+\kappa_{2,x}+\kappa_{3,x}+\kappa_{1,z}+\kappa_{2,z}+\kappa_{3,z})t}
\nonumber \\ &&
\quad (e^{\kappa_{1,x}t}- e^{\kappa_{2,x}t}-e^{\kappa_{3,x}t}
+e^{(\kappa_{1,x}+\kappa_{2,x}+\kappa_{3,x})t}) \nonumber \\
\end{eqnarray}

Under the simultaneous action of all the NMR noise channels, the W
state decoheres as:
\begin{equation}
\rho_{W} = \left(
\begin{array}{cccccccc}
 \alpha_{1}  & 0 & 0 & \beta_{1} & 0 & \beta_{5} & \beta_{1} &0 \\
 0 & \alpha_{2}  & \beta_{2} & 0 & \beta_{6} & 0 & 0& \beta_{10} \\
 0 & \beta_{2} & \alpha_{3}  & 0 & \beta_{11}& 0 & 0 & \beta_{7} \\
 \beta_{1} & 0 & 0 & \alpha_{4}  & 0 & \beta_{12} & \beta_{8} & 0 \\
 0 & \beta_{6} & \beta_{11} & 0 & \alpha_{5}  & 0 & 0 & \beta_{3} \\
 \beta_{5} & 0 & 0 & \beta_{12} & 0 & \alpha_{6}  & \beta_{4} & 0 \\
 \beta_{1} & 0 & 0 &\beta_{8} & 0 & \beta_{4} & \alpha_{7}  & 0 \\
0 & \beta_{10}&\beta_{7} & 0 & \beta_{3} & 0 & 0 & \alpha_{8} \\
\end{array}
\right) \nonumber \\
\end{equation}
Where
\begin{eqnarray}
\alpha_{1}&=&\frac{1}{8}- \frac{1}{24} e^{-(\kappa_{x,1}+\kappa_{x,2}+\kappa_{x,3})t} (3+e^{\kappa_{x,1}t} 
 +e^{\kappa_{x,2}t}- 
\nonumber \\ &&
e^{(\kappa_{x,1}+\kappa_{x,2})t}+e^{\kappa_{x,3}t} 
 - e^{(\kappa_{x,1}+\kappa_{x,3})t} - e^{(\kappa_{x,2}+\kappa_{x,3})t}) \nonumber \\
\alpha_{2}&=&\frac{1}{8}+ \frac{1}{24} e^{-(\kappa_{x,1}+\kappa_{x,2}+\kappa_{x,3})t} (3+e^{\kappa_{x,1}t} 
+e^{\kappa_{x,2}t}-
\nonumber \\ && 
e^{(\kappa_{x,1}+\kappa_{x,2})t}-e^{\kappa_{x,3}t}
+ e^{(\kappa_{x,1}+\kappa_{x,3})t} +
e^{(\kappa_{x,2}+\kappa_{x,3})t}) \nonumber \\ 
\alpha_{3}&=&\frac{1}{8}+ \frac{1}{24}
e^{-(\kappa_{x,1}+\kappa_{x,2}+\kappa_{x,3})t}
(3+e^{\kappa_{x,1}t} 
-e^{\kappa_{x,2}t}
\nonumber \\ && 
+ e^{(\kappa_{x,1}+\kappa_{x,2})t}+e^{\kappa_{x,3}t} 
- e^{(\kappa_{x,1}+\kappa_{x,3})t} + e^{(\kappa_{x,2}+\kappa_{x,3})t}) \nonumber \\
\alpha_{4}&=&\frac{1}{8}- \frac{1}{24}
e^{-(\kappa_{x,1}+\kappa_{x,2}+\kappa_{x,3})t}
(3+e^{\kappa_{x,1}t} 
 -e^{\kappa_{x,2}t}
\nonumber \\ && 
+ e^{(\kappa_{x,1}+\kappa_{x,2})t}-e^{\kappa_{x,3}t}
+ e^{(\kappa_{x,1}+\kappa_{x,3})t} - e^{(\kappa_{x,2}+\kappa_{x,3})t}) 
\nonumber \\
\alpha_{5}&=&\frac{1}{8}+ \frac{1}{24} e^{-(\kappa_{x,1}+\kappa_{x,2}+\kappa_{x,3})t} (3-e^{\kappa_{x,1}t} 
+e^{\kappa_{x,2}t}
\nonumber \\ && 
+ e^{(\kappa_{x,1}+\kappa_{x,2})t}+e^{\kappa_{x,3}t} 
+ e^{(\kappa_{x,1}+\kappa_{x,3})t} - e^{(\kappa_{x,2}+\kappa_{x,3})t}) \nonumber \\
\alpha_{6}&=&\frac{1}{8}+ \frac{1}{24} e^{-(\kappa_{x,1}+\kappa_{x,2}+\kappa_{x,3})t} (-3+e^{\kappa_{x,1}t} 
-e^{\kappa_{x,2}t}
\nonumber \\ && 
- e^{(\kappa_{x,1}+\kappa_{x,2})t}+e^{\kappa_{x,3}t} 
+ e^{(\kappa_{x,1}+\kappa_{x,3})t} - e^{(\kappa_{x,2}+\kappa_{x,3})t}) \nonumber \\
\alpha_{7}&=&\frac{1}{8}+ \frac{1}{24} e^{-(\kappa_{x,1}+\kappa_{x,2}+\kappa_{x,3})t} (-3+e^{\kappa_{x,1}t}+ 
e^{\kappa_{x,2}t}
\nonumber \\ &&
+ e^{(\kappa_{x,1}+\kappa_{x,2})t}-e^{\kappa_{x,3}t} 
- e^{(\kappa_{x,1}+\kappa_{x,3})t} - e^{(\kappa_{x,2}+\kappa_{x,3})t}) \nonumber \\
\alpha_{8}&=&\frac{1}{8}- \frac{1}{24} e^{-(\kappa_{x,1}+\kappa_{x,2}+\kappa_{x,3})t} (-3+e^{\kappa_{x,1}t}+ 
e^{\kappa_{x,2}t}
\nonumber \\ &&
+ e^{(\kappa_{x,1}+\kappa_{x,2})t}+e^{\kappa_{x,3}t} 
+ e^{(\kappa_{x,1}+\kappa_{x,3})t} + e^{(\kappa_{x,2}+\kappa_{x,3})t}) \nonumber 
\end{eqnarray}
\begin{eqnarray}
\beta_{1}&=&\frac{1}{12} (e^{-(\kappa_{x,1}+\kappa_{x,2}+\kappa_{x,3}+\kappa_{z,2}+\kappa_{z,3})t}\nonumber \\
&&(1+e^{(\kappa_{x,1})t})(-1+e^{(\kappa_{x,2}+\kappa_{x,3})t}))\nonumber \\
\beta_{2}&=&\frac{1}{12} (e^{-(\kappa_{x,1}+\kappa_{x,2}+\kappa_{x,3}+\kappa_{z,2}+\kappa_{z,3})t}\nonumber \\
&&(1+e^{(\kappa_{x,1})t})(1+e^{(\kappa_{x,2}+\kappa_{x,3})t}))\nonumber \\
\beta_{3}&=&\frac{1}{12} (e^{-(\kappa_{x,1}+\kappa_{x,2}+\kappa_{x,3}+\kappa_{z,2}+\kappa_{z,3})t}\nonumber \\
&&(-1+e^{(\kappa_{x,1})t})(-1+e^{(\kappa_{x,2}+\kappa_{x,3})t}))\nonumber \\
\beta_{4}&=&\frac{1}{12} (e^{-(\kappa_{x,1}+\kappa_{x,2}+\kappa_{x,3}+\kappa_{z,2}+\kappa_{z,3})t}\nonumber \\
&&(-1+e^{(\kappa_{x,1})t})(1+e^{(\kappa_{x,2}+\kappa_{x,3})t}))\nonumber 
\end{eqnarray}
\begin{eqnarray}
\beta_{5}&=&\frac{1}{12} (e^{-(\kappa_{x,1}+\kappa_{x,2}+\kappa_{x,3}+\kappa_{z,1}+\kappa_{z,3})t}\nonumber \\
&&(1+e^{(\kappa_{x,2})t})(-1+e^{(\kappa_{x,1}+\kappa_{x,3})t}))\nonumber \\
\beta_{6}&=&\frac{1}{12} (e^{-(\kappa_{x,1}+\kappa_{x,2}+\kappa_{x,3}+\kappa_{z,1}+\kappa_{z,3})t}\nonumber \\
&&(1+e^{(\kappa_{x,2})t})(1+e^{(\kappa_{x,1}+\kappa_{x,3})t}))\nonumber \\
\beta_{7}&=&\frac{1}{12} (e^{-(\kappa_{x,1}+\kappa_{x,2}+\kappa_{x,3}+\kappa_{z,1}+\kappa_{z,3})t}\nonumber \\
&&(-1+e^{(\kappa_{x,2})t})(-1+e^{(\kappa_{x,1}+\kappa_{x,3})t}))\nonumber \\
\beta_{8}&=&\frac{1}{12} (e^{-(\kappa_{x,1}+\kappa_{x,2}+\kappa_{x,3}+\kappa_{z,1}+\kappa_{z,3})t}\nonumber \\
&&(-1+e^{(\kappa_{x,2})t})(1+e^{(\kappa_{x,1}+\kappa_{x,3})t}))\nonumber \\
\beta_{9}&=&\frac{1}{12} (e^{-(\kappa_{x,1}+\kappa_{x,2}+\kappa_{x,3}+\kappa_{z,1}+\kappa_{z,2})t}\nonumber \\
&&(-1+e^{(\kappa_{x,1}+\kappa_{x,2})t})(1+e^{(\kappa_{x,3})t}))\nonumber \\
\beta_{10}&=&\frac{1}{12} (e^{-(\kappa_{x,1}+\kappa_{x,2}+\kappa_{x,3}+\kappa_{z,1}+\kappa_{z,2})t}\nonumber \\
&&(-1+e^{(\kappa_{x,1}+\kappa_{x,2})t})(-1+e^{(\kappa_{x,3})t}))\nonumber \\
\beta_{11}&=&\frac{1}{12} (e^{-(\kappa_{x,1}+\kappa_{x,2}+\kappa_{x,3}+\kappa_{z,1}+\kappa_{z,2})t}\nonumber \\
&&(1+e^{(\kappa_{x,1}+\kappa_{x,2})t})(1+e^{(\kappa_{x,3})t}))\nonumber \\
\beta_{12}&=&\frac{1}{12} (e^{-(\kappa_{x,1}+\kappa_{x,2}+\kappa_{x,3}+\kappa_{z,1}+\kappa_{z,2})t}\nonumber \\
&&(1+e^{(\kappa_{x,1}+\kappa_{x,2})t})(-1+e^{(\kappa_{x,3})t}))
\end{eqnarray}

Under the simultaneous action of all the 
NMR noise channels, the ${\rm W \bar{W}}$
state decoheres as:
\begin{equation}
\rho_{{\rm W \bar{W}}} =\left(
\begin{array}{cccccccc}
 \alpha_{1}  & \beta_{1} & \beta_{2} & \beta_{3} &\beta_{4} & \beta_{5} & \beta_{6} &\beta_{7} \\
  \beta_{1}  & \alpha_{2}  & \beta_{8} &\beta_{9} & \beta_{10} & \beta_{11} & \beta_{12} & \beta_{13} \\
  \beta_{2}  & \beta_{8} & \alpha_{3}  &  \beta_{14}  & \beta_{15}&  \beta_{16}  &  \beta_{11} & \beta_{5} \\
 \beta_{3} &  \beta_{9}  & \beta_{14}  & \alpha_{4}  &  \beta_{17}  & \beta_{15} & \beta_{10} &  \beta_{4} \\
  \beta_{4}  & \beta_{10} & \beta_{15} &  \beta_{17}  & \alpha_{4}  & \beta_{15}  &  \beta_{9}  & \beta_{18} \\
 \beta_{5} &  \beta_{11}  &  \beta_{16}  & \beta_{15} & \beta_{15}  & \alpha_{3}  & \beta_{8} &  \beta_{2}  \\
 \beta_{6} &  \beta_{12}  &  \beta_{11} &\beta_{10} &  \beta_{9} & \beta_{8} & \alpha_{2}  &  \beta_{1}  \\
 \beta_{7}  & \beta_{13}&\beta_{5} &  \beta_{4}  & \beta_{18} &  \beta_{2} &  \beta_{1} & \alpha_{1} \\
\end{array}
\right) 
\end{equation}
where
\begin{eqnarray}
\alpha_{1}&=&\frac{1}{24}(3-e^{-(\kappa_{x,1}+\kappa_{x,2})t}-e^{-(\kappa_{x,1}+\kappa_{x,3})t} \nonumber \\
&&- e^{-(\kappa_{x,2}+\kappa_{x,3})t})  \nonumber \\
\alpha_{2}&=&\frac{1}{24}(3-e^{-(\kappa_{x,1}+\kappa_{x,2})t}+e^{-(\kappa_{x,1}+\kappa_{x,3})t} \nonumber \\
&& + e^{-(\kappa_{x,2}+\kappa_{x,3})t})  \nonumber \\
\alpha_{3}&=&\frac{1}{24}(3+e^{-(\kappa_{x,1}+\kappa_{x,2})t}-e^{-(\kappa_{x,1}+\kappa_{x,3})t} \nonumber \\
&& + e^{-(\kappa_{x,2}+\kappa_{x,3})t})  \nonumber \\
\alpha_{4}&=&\frac{1}{24}(3+e^{-(\kappa_{x,1}+\kappa_{x,2})t}+e^{-(\kappa_{x,1}+\kappa_{x,3})t} \nonumber \\
&& - e^{-(\kappa_{x,2}+\kappa_{x,3})t})  \nonumber \\
\beta_{1}&=&\frac{1}{12} e^{-(\kappa_{x,1}+\kappa_{x,2}+2\kappa_{z,3})t} \nonumber \\
&&(e^{(\kappa_{x,1}+\kappa_{x,2}+\kappa_{z,3})t}-e^{ \kappa_{z,3} t}) \nonumber \\
\beta_{2}&=&\frac{1}{12} e^{-(\kappa_{x,1}+\kappa_{x,3}+2\kappa_{z,2})t} \nonumber \\
&&(e^{(\kappa_{x,1}+\kappa_{x,3}+\kappa_{z,2})t}-e^{ \kappa_{z,2} t}) \nonumber \\
\beta_{3}&=&\frac{1}{12} e^{-(\kappa_{x,2}+\kappa_{x,3}+2(\kappa_{z,2}+\kappa_{z,3}))t} \nonumber \\
&&(e^{(\kappa_{x,2}+\kappa_{x,3}+\kappa_{z,2}+\kappa_{z,3})t}-e^{ (\kappa_{z,2}+\kappa_{z,3}) t}) \nonumber \\
\beta_{4}&=&\frac{1}{12} e^{-(\kappa_{x,2}+\kappa_{x,3}+2\kappa_{z,1})t} \nonumber \\
&&(-e^{ \kappa_{z,1} t}+e^{(\kappa_{x,2}+\kappa_{x,3}+\kappa_{z,1})t}) \nonumber \\
\beta_{5}&=&\frac{1}{12} e^{-(\kappa_{x,1}+\kappa_{x,3}+2(\kappa_{z,1}+\kappa_{z,3}))t} \nonumber \\
&&(-e^{ (\kappa_{z,1}+\kappa_{z,3}) t}+e^{(\kappa_{x,1}+\kappa_{x,3}+\kappa_{z,1}+\kappa_{z,3})t}) \nonumber \\
\beta_{6}&=&\frac{1}{12} e^{-(\kappa_{x,1}+\kappa_{x,2}+2(\kappa_{z,1}+\kappa_{z,2}))t} \nonumber \\
&&(-e^{ (\kappa_{z,1}+\kappa_{z,2}) t}+e^{(\kappa_{x,1}+\kappa_{x,2}+\kappa_{z,1}+\kappa_{z,2})t}) \nonumber \\
\beta_{7}&=&-\frac{1}{24} e^{-(\kappa_{x,1}+\kappa_{x,2}+\kappa_{x,3}+\kappa_{z,1}+\kappa_{z,2}+\kappa_{z,3})t} \nonumber \\
&& (e^{\kappa_{x,1} t}+e^{\kappa_{x,2} t}+e^{\kappa_{x,3} t}-3e^{(\kappa_{x,1}+\kappa_{x,2}+\kappa_{x,3})t}) \nonumber \\
\beta_{8}&=&\frac{1}{12} e^{-(\kappa_{x,2}+\kappa_{x,3}+2(\kappa_{z,2}+\kappa_{z,3}))t} \nonumber \\
&&(e^{ (\kappa_{z,2}+\kappa_{z,3}) t}+e^{(\kappa_{x,2}+\kappa_{x,3}+\kappa_{z,2}+\kappa_{z,3})t}) \nonumber \\
\beta_{9}&=&\frac{1}{12} e^{-(\kappa_{x,1}+\kappa_{x,3}+2\kappa_{z,2})t} \nonumber \\
&&(e^{ \kappa_{z,2} t}+e^{(\kappa_{x,1}+\kappa_{x,3}+\kappa_{z,2})t}) \nonumber \\
\beta_{10}&=&\frac{1}{12} e^{-(\kappa_{x,1}+\kappa_{x,3}+2(\kappa_{z,1}+\kappa_{z,3}))t} \nonumber \\
&&(e^{ (\kappa_{z,1}+\kappa_{z,3}) t}+e^{(\kappa_{x,1}+\kappa_{x,3}+\kappa_{z,1}+\kappa_{z,3})t}) \nonumber 
\end{eqnarray}

\begin{eqnarray}
\beta_{11}&=&\frac{1}{12} e^{-(\kappa_{x,2}+\kappa_{x,3}+2\kappa_{z,1})t}(e^{ \kappa_{z,1} t}+e^{(\kappa_{x,2}+\kappa_{x,3}+\kappa_{z,1})t}) \nonumber \\
\beta_{12}&=&\frac{1}{24} e^{-(\kappa_{x,1}+\kappa_{x,2}+\kappa_{x,3}+\kappa_{z,1}+\kappa_{z,2}+\kappa_{z,3})t} \nonumber \\
&& (e^{\kappa_{x,1} t}+e^{\kappa_{x,2} t}-e^{\kappa_{x,3} t}+3e^{(\kappa_{x,1}+\kappa_{x,2}+\kappa_{x,3})t}) \nonumber \\
\beta_{13}&=&\frac{1}{12} e^{-(\kappa_{x,1}+\kappa_{x,2}+2(\kappa_{z,1}+\kappa_{z,2}))t} \nonumber \\
&&(-e^{ (\kappa_{z,1}+\kappa_{z,2}) t}+e^{(\kappa_{x,1}+\kappa_{x,2}+\kappa_{z,1}+\kappa_{z,2})t}) \nonumber \\
\beta_{14}&=&\frac{1}{12} e^{-(\kappa_{x,1}+\kappa_{x,2}+2\kappa_{z,3})t} \nonumber \\
&&(e^{ \kappa_{z,3} t}+e^{(\kappa_{x,1}+\kappa_{x,2}+\kappa_{z,3})t}) \nonumber \\
\beta_{15}&=&\frac{1}{12} e^{-(\kappa_{x,1}+\kappa_{x,2}+2(\kappa_{z,1}+\kappa_{z,2}))t} \nonumber \\
&&(e^{ (\kappa_{z,1}+\kappa_{z,2}) t}+e^{(\kappa_{x,1}+\kappa_{x,2}+\kappa_{z,1}+\kappa_{z,2})t}) \nonumber \\
\beta_{16}&=&\frac{1}{24} e^{-(\kappa_{x,1}+\kappa_{x,2}+\kappa_{x,3}+\kappa_{z,1}+\kappa_{z,2}+\kappa_{z,3})t} \nonumber \\
&& (e^{\kappa_{x,1} t}-e^{\kappa_{x,2} t}+e^{\kappa_{x,3} t}+3e^{(\kappa_{x,1}+\kappa_{x,2}+\kappa_{x,3})t}) \nonumber \\
\beta_{17}&=&\frac{1}{24} e^{-(\kappa_{x,1}+\kappa_{x,2}+\kappa_{x,3}+\kappa_{z,1}+\kappa_{z,2}+\kappa_{z,3})t} \nonumber \\
&& (-e^{\kappa_{x,1} t}+e^{\kappa_{x,2} t}+e^{\kappa_{x,3} t}+3e^{(\kappa_{x,1}+\kappa_{x,2}+\kappa_{x,3})t}) \nonumber \\
\beta_{18}&=&\frac{1}{12} e^{-(\kappa_{x,2}+\kappa_{x,3}+2(\kappa_{z,2}+\kappa_{z,3}))t} \nonumber \\
&&(e^{ (\kappa_{z,2}+\kappa_{z,3}) t}+e^{(\kappa_{x,2}+\kappa_{x,3}+\kappa_{z,2}+\kappa_{z,3})t}) 
\end{eqnarray}

Solving the master equation (Eqn.~\ref{mastereqn}) ensures that the
off-diagonal elements of the corresponding $\rho$ matrices satisfy
a set of coupled equations, from which the explicit values of 
$\alpha$s and $\beta$s can be computed. The equations are solved in
the high-temperature limit. For an ensemble of NMR spins at room temperature 
this implies that the energy $E << k_{B} T$ where $k_B$ is the Boltzmann
constant and $T$ refers to the temperature, ensuring a Boltzmann
distribution of spin populations at thermal equilibrium.
%

\end{document}